\documentclass[nofootinbib,twocolumn]{revtex4}

\begin{document}

\title{On dark degeneracy and interacting models}
\author{S. Carneiro\footnote{saulo.carneiro@pq.cnpq.br} and H. A. Borges\footnote{humberto@ufba.br}}

\affiliation{Instituto de F\'{\i}sica, Universidade Federal da Bahia\\ Campus de Ondina, Salvador, BA, Brazil, 40210-340}


\begin{abstract}
Cosmological background observations cannot fix the dark energy equation of state, which is related to a degeneracy in the definition of the dark sector components. Here we show that this degeneracy can be broken at perturbation level by imposing two observational properties on dark matter. First, dark matter is defined as the clustering component we observe in large scale structures. This definition is meaningful only if dark energy is unperturbed, which is achieved if we additionally assume, as a second condition, that dark matter is cold, i.e. non-relativistic. As a consequence, dark energy models with equation-of-state parameter $-1 \le\omega< 0$ are reduced to two observationally distinguishable classes with $\omega = -1$, equally competitive when tested against observations. The first comprises the $\Lambda$CDM model with constant dark energy density. The second consists of interacting models with an energy flux from dark energy to dark matter.
\end{abstract}

\maketitle

\section{Introduction}
\label{Introduction}

The derivation of vacuum energy in expanding spacetime is probably the most difficult challenge in the interface of theoretical cosmology and quantum field theories. Sometimes this problem is posed as a huge difference between the vacuum energy density, regularised in some way, and the observed value of the cosmological term. If the divergent vacuum density is regularised by imposing an ultraviolet cutoff of the order of the Planck mass, this results in a theoretical value $122$ orders of magnitude above the observed value. Even if we use for the cutoff the lowest energy scale available, given by the QCD vacuum transition, the result is still $40$ orders above the observed one. The problem is, however, more subtle. Einstein's equations tell us that in flat spacetime the vacuum energy-momentum tensor is zero. Therefore, the vacuum density in curved spacetimes must be derived by an appropriate renormalisation procedure which subtracts the divergent contribution in flat spacetime. For conformal fields in de Sitter spacetime this leads to a vacuum density $\Lambda \approx H^4$, where $H$ is the expansion rate \cite{Ford}. Now this gives a tiny value for the present vacuum density, many orders below the observed cosmological term. An alternative may be to assume that vacuum somehow does not gravitate and that the cosmological constant is a geometrical term added {\it ad hoc} in Einstein's equations, with no microphysics determining its scale.

In the high energy limit, the result $\Lambda \approx H^4$ can be used to obtain a non-singular scenario where a radiation phase is born from a previous de Sitter phase through a continuous and fast transition along which the vacuum density decays producing relativistic matter \cite{H4}. However, any model with matter creation from a decaying vacuum has phenomenological status because vacuum energy-momentum conservation is one of the conditions behind renormalisation techniques \cite{Parker}. In \cite{Inflation} it is argued that the backreaction of relativistic particles creation leads to a vacuum density scaling with $H^2$, which in turn gives rise to an inflationary period with desired features. Another context where these results are not valid is the case of interacting fields. Estimates of the energy density of the QCD vacuum condensate in the low energy FLRW spacetime leads to $\Lambda \approx m^3 H$, where $m \approx 150$ MeV is the energy scale of the QCD phase transition \cite{Schutz}. By taking an approximate de Sitter universe with $\Lambda \approx H^2$, this leads to $\Lambda \approx m^6$, which coincides with the observed value of $\Lambda$. Needless to say, these estimates are not free of the difficulties inherent to non-perturbative QCD in curved backgrounds.

These theoretical limitations oblige us to look for an observational route to the late-time cosmological term, usually referred to as dark energy (DE). In a de Sitter spacetime, its equation of state would be necessarily given by $p_{\Lambda} = -\Lambda$, owing to the symmetry of this background. In the present FLRW spacetime, observations show that the Universe is in accelerated expansion. This means that the dark energy equation-of-state parameter $\omega$, defined by
\begin{equation} \label{3}
p = \omega \rho,
\end{equation}
is a negative function of time, to be determined\footnote{We are assuming that the dark fluid respects the weak energy condition, i.e. $\rho > 0$ and $-1 \le \omega < 0$. The {\it phantom} case $\omega<-1$ will not be considered here.}. However, as pointed out in \cite{Kunz}, background observations cannot uniquely fix this function because the effects of DE and dark matter (DM) are degenerated in a way that will be clarified below.  In this letter we will show, in Section \ref{Dark degeneracy}, that such a degeneracy can be broken at perturbation level, provided that:
\begin{itemize}
\item{Dark matter is defined as the clustering component observed in galaxies and clusters.}
\item{Dark matter is assumed to be cold, that is, non-relativistic.}
\end{itemize}
On this basis the degeneracy will be reduced to two distinct classes of DE models with $\omega = -1$, namely the $\Lambda$CDM model, with a constant $\Lambda$, and interacting models with an energy flux from DE to DM. In Section \ref{Breaking the degeneracy} we will consider the particular case of a constant DM creation rate, which corresponds to $\Lambda \propto H$. The resulting model is not reducible to the $\Lambda$CDM model. Both models have the same number of free parameters and are equally competitive when tested against observations. In Section \ref{Examples} we will see two specific examples of dark degeneracy, namely the (non-adiabatic) generalised Chaplygin gas and the quintessence field, showing how they realise the interacting model.

\section{Dark degeneracy}
\label{Dark degeneracy}

In the present letter dark degeneracy is formulated in the following manner. Let us split dark fluid (\ref{3}) as
\begin{eqnarray}
\rho &=& \Lambda + \rho_m, \label{4} \\
p_{\Lambda} &=& - \Lambda, \label{5} \\
p_m &=& \omega_m \rho_m, \label{6}
\end{eqnarray}
with $\omega_m \ge 0$ and $\Lambda > 0$. From (\ref{3})-(\ref{6}) we have
\begin{equation} \label{7}
\rho_m = \frac{\omega + 1}{\omega_m - \omega} \Lambda.
\end{equation}
For $-1 \le \omega <0$ we have $\rho_m \ge 0$, and this component can be interpreted as dark matter. This degeneracy in the definitions of DE and DM is unavoidable at the background level, and two illustrative examples will be given in Section \ref{Examples}. Nevertheless, we will see below that the degeneracy can be broken at perturbation level if we correctly define the observed dark matter and assume that DM is cold, i.e. non-relativistic.

\subsection{Background}
\label{Background}

With densities and pressures given by (\ref{4})-(\ref{6}) (from now on we will take $\omega_m=0$), the Friedmann and continuity equations assume the form\footnote{We are considering the spatially flat case and using $c = \hbar = 8\pi G = 1$.}
\begin{eqnarray}
\rho_m + \Lambda &=& 3 H^2, \label{9}\\
\dot{\rho}_m + 3 H \rho_m &=& - \dot{\Lambda}, \label{8} 
\end{eqnarray}
where the dot means derivative with respect to cosmological time. Equation (\ref{8}) expresses total energy conservation, and it means that the two components may interact in general \cite{Chimento}. Differentiating (\ref{9}) and using (\ref{8}) we derive
\begin{equation}\label{10}
\rho_m = - 2 \dot{H}.
\end{equation}
Using (\ref{7}) in (\ref{9}) we have
\begin{equation}\label{11}
\Lambda = -3\omega H^2.
\end{equation}
Differentiating (\ref{11}) and using (\ref{10}), (\ref{9}) and (\ref{7}) we obtain
\begin{equation}\label{12}
\dot{\Lambda} = \left( 3\omega H - \frac{\dot{\omega}}{\omega+1} \right) \rho_m.
\end{equation}
Hence, equation (\ref{8}) can be rewritten as
\begin{equation}\label{13}
\dot{\rho}_m + 3 H \rho_m = \Gamma \rho_m,
\end{equation}
with $\Gamma$ given by
\begin{equation}\label{14}
\Gamma = \frac{\dot{\omega}}{\omega+1} - 3\omega H.
\end{equation}
Writing $\rho_m = Mn$, where $M$ and $n$ are defined as the DM particle mass and number density respectively, and using $H = \dot{a}/a$, where $a$ is the scale factor, we obtain from (\ref{13})
\begin{equation}
\frac{1}{a^3}\frac{d}{dt}(a^3n) = \Gamma n,
\end{equation}
which shows that $\Gamma$ defines the rate of DM creation.

\subsection{Perturbations}
\label{Perturbations}

The energy-conservation equation (\ref{8}) can be written in the covariant form
\begin{eqnarray}
T_{m\,;\nu}^{\mu\nu} &=& Q^{\mu}, \label{15} \\
T_{\Lambda\,;\nu}^{\mu\nu} &=& - Q^{\mu}, \label{16} 
\end{eqnarray}
where $T_A^{\mu\nu}$ ($A = m,\Lambda$) are the energy-momentum tensors of each component, and $Q^{\mu}$ is the energy-momentum transfer between DE and DM. The latter can be decomposed as
\begin{equation} \label{17}
Q^{\mu} = Q u^{\mu} + \bar{Q}^{\mu},
\end{equation}
with $u^{\mu}\bar{Q}_{\mu} = 0$, where $u^{\mu}$ is the dark fluid $4$-velocity. From (\ref{5}) we see that $T_{\Lambda}^{\mu\nu} = \Lambda g^{\mu\nu}$, which shows that $\Lambda$ is an invariant. Hence, from (\ref{16}) it is easy to obtain
\begin{eqnarray}
Q &=& - \Lambda_{,\nu} u^{\nu}, \\
\bar{Q}^{\mu} &=& \Lambda_{,\nu} \left( u^{\mu} u^{\nu} - g^{\mu\nu} \right) \label{barQ}.
\end{eqnarray}
In the dark fluid comoving frame we have $\bar{Q}^{\mu} = 0$, and $Q = -\dot{\Lambda} = \Gamma \rho_m$. The former shows that there is no momentum transfer in the isotropic background, while the latter shows that $Q$ represents the energy transfer between the components. For pressureless DM we can write $Q = \Gamma g_{\mu\nu} T_{m}^{\mu\nu}$,
which shows that $\Gamma$ is also invariant. 

A linear perturbation of (\ref{barQ}) leads to
\begin{eqnarray}
\delta \bar{Q}^0 &=& 0, \\
\delta \bar{Q}_{i} &=& (\delta \Lambda + \dot{\Lambda} \theta)_{,i}, \label{humberto}
\end{eqnarray}
where $\theta$ is the dark fluid velocity potential, $\vec{\nabla}\theta = \delta \vec{u}$. Assuming that dark matter is non-relativistic, the momentum transfer should be negligible compared to the energy transfer, that is, dark matter particles are created with negligible momentum compared to their rest energy. This means that $\bar{Q}^{\mu}$ is zero at both background and perturbation levels, which also means that DM follows geodesics in the dark fluid rest frame \cite{Maartens,Wands}. Hence, from (\ref{humberto}) we obtain
\begin{equation} \label{20}
\delta \Lambda^c \equiv \delta \Lambda + \dot{\Lambda} \theta = 0,
\end{equation}
where $\delta \Lambda^c$ is the (gauge-invariant) comoving perturbation of the DE component \cite{Karim}. Equation (\ref{20}) means that if dark matter in the decomposition (\ref{4}) is non-relativistic, the dark energy defined by (\ref{4}) and (\ref{5}) does not cluster. Therefore, the DM component defined in (\ref{4}) and (\ref{6}) coincides with that observed in galaxies and clusters.

Condition (\ref{20}) is also necessary to avoid the presence of unobserved oscillations and instabilities in the DM power spectrum. From the perturbed Einstein equations for a perfect fluid we have \cite{GCG}
\begin{eqnarray} \label{tam} 
\Phi_{B}''+3\mathcal{H}(1+c_{a}^2)\Phi_{B}'&+& \\ \nonumber
[2\mathcal{H}'+(1+3c_{a}^2)\mathcal{H}^2+c_{s}^2k^2]\Phi_B &=& 0\,,
\end{eqnarray}
where $\Phi_B$ is the Bardeen gravitational potential, $\mathcal{H}=aH$, $k$ is the perturbation comoving wave-number, $c_a^2 = \dot{p}/\dot{\rho}$ is the squared adiabatic sound speed, $c_s^2 = \delta p^c/\delta \rho^c$ is the squared effective sound speed, and the prime means derivative with respect to conformal time. As in (\ref{20}), the uper-index $c$ means a gauge-invariant comoving perturbation. The term proportional to $k^2$ in (\ref{tam}) induces oscillations and instabilities in the gravitational potential, which are reflected in the matter power spectrum through the Poisson equation. In order to avoid these oscillations and instabilities we must have $|c_s| \ll 1$. For the two-component fluid (\ref{4}) we have $c_s^2 = -\delta \Lambda^c/\delta \rho^c$, and the last condition is satisfied for $\delta \Lambda^c = 0$, that is, $\bar{Q}^{\mu} = 0$. As discussed in \cite{GCG}, $c_s^2 \neq c_a^2$ owing to the presence of late-time entropic perturbations related to the energy flux between dark energy and dark matter.

The results of this section can be summarised in the following theorem and corollaries:

${ }$

\noindent{\it Theorem}: Let it be a unified dark fluid DF with equation of state $p =\omega \rho$, with $-1 \le \omega < 0$. (i) DF is observationally indistinguishable from a fluid formed by pressureless dark matter DM and a dark energy DE with equation-of-state parameter $\omega = -1$. In general these components interact and DF is non-adiabatic. (ii) If the momentum transfer between the components is negligible, DE does not cluster and DM coincides with clustering matter. This is the case if DM is non-relativistic. 

${ }$

\noindent {\it Corollaries}: Let it be a dark energy candidate with equation-of-state parameter $-1 \le \omega < 0$. (i) Any observational analysis which identifies cold dark matter with clustering matter leads to $\omega \approx -1$. (ii) If, in addition, such an analysis assumes that dark matter is conserved, the concordance is given by the $\Lambda$CDM model. 

\section{Breaking degeneracy}
\label{Breaking the degeneracy}

In the spatially flat FLRW spacetime there are two physical quantities that can be naturally identified as the creation rate $\Gamma$ (apart from the trivial case $\Gamma = 0$, for which we have conserved DM and a constant DE density, that is, a cosmological constant). Namely, the expansion rate $H$ and the DM particle mass, which determines its quantum fluctuation rate.  In \cite{Inflation} it was shown that in the early-time limit of relativistic particles the case $\Gamma \propto H$ leads to $\Lambda \propto H^2$ and to a viable inflationary scenario.  In the late-time limit as well, by taking $\Gamma = \gamma H$, equations (\ref{13}), (\ref{10}) and (\ref{8}) are compatible only if
\begin{equation} \label{21}
\Lambda = \gamma H^2 + \lambda,
\end{equation}
where $\lambda$ is an integration constant. If we choose $\lambda= 0$, the resulting cosmology is not in accordance with observations because, from (\ref{9}), both $\rho_m$ and $\Lambda$ will scale as $H^2$ and there will be no transition from a decelerated phase to an accelerated one\footnote{Adding conserved baryons with $\Omega_b \approx 0.05$ would not give the correct transition redshift.}. On the other hand, if we take $\lambda \neq 0$ \cite{Waga}, observations will favor $\gamma \approx 0$, that is, the $\Lambda$CDM model, since the latter fits observations very well.

Now, let us consider the second possibility, a constant $\Gamma$. The consistency between (\ref{13}), (\ref{10}) and (\ref{8}) is only possible if
\begin{equation} \label{22}
\Lambda = 2\Gamma H +\lambda.
\end{equation}
For $\Gamma = 0$ we have no DM creation and a constant $\Lambda$. Since there is no physical scale fixing the integration constant $\lambda$ (the cosmological constant problem), let us take it as zero and
\begin{equation} \label{22'}
\Lambda = 2 \Gamma H.
\end{equation}
Equations (\ref{11}), (\ref{14}) and (\ref{22'}) lead to the evolution equation
\begin{equation} \label{23'}
3H^2 - 2\Gamma H +2\dot{H} = 0.
\end{equation}
The solution is given by \cite{Borges}
\begin{equation} \label{Borges}
H = \frac{2\Gamma/3}{1 - e^{-\Gamma t}},
\end{equation}
where an integration constant was chosen so that $H \rightarrow \infty$ when $t \rightarrow 0$. It represents a universe that evolves from an Einstein-de Sitter phase, dominated by matter, to an asymptotically de Sitter era. This model has no free parameters apart from the standard ones, namely the present values of the Hubble parameter and of the matter relative density. It is not reducible to the $\Lambda$CDM model in any limit and has been shown to be competitive when compared to the latter. It has been tested against the most precise observations, with very good concordance \cite{tests}. Furthermore, by writing (\ref{22'}) in the covariant form $\Lambda = 2\Gamma \Theta/3$, where $\Theta = u^{\nu}_{;\nu}$ is the expansion scalar, it was explicitly shown in \cite{Zimdahl} that $\delta \Lambda^c \ll \delta \rho_m^c$ in the observed scales, as required in order to have a coincidence between our definition of DM and that observed in galaxies and clusters. From (\ref{Borges}) we see that $\Gamma = 3H_{dS}/2$, where $H_{dS}$ is the expansion rate in the de Sitter limit.

\section{Examples}
\label{Examples}

Let us now examine two known examples of dark energy models where dark degeneracy can be made explicit. In both examples we will show how to realise the particular case of a constant $\Gamma$.

\subsection{Non-adiabatic Chaplygin gas}
\label{Non-adiabatic Chaplygin gas}

The generalised Chaplygin gas (GCG) is characterised by an adiabatic sound speed \cite{Wands,GCG1,GCG}
\begin{equation}\label{24}
c_a^2 = \frac{\dot{p}}{\dot{\rho}} = - \alpha \omega,
\end{equation}
where $\alpha$ is a constant. Hence, by differentiating $p = \omega \rho$ we obtain
\begin{equation}\label{25}
\dot{\omega}\rho = -\omega (\alpha + 1) \dot{\rho}.
\end{equation}
Using the conservation equation
\begin{equation}\label{26}
\dot{\rho} + 3H (\rho + p) = 0,
\end{equation}
it follows that
\begin{equation}\label{27}
\dot{\omega} = 3 \omega (\alpha + 1) (\omega + 1) H.
\end{equation}

The gas can be split into two interacting components like in (\ref{4})-(\ref{6}). Substituting (\ref{27}) in (\ref{14}) we obtain the rate of matter creation,
\begin{equation}\label{28}
\Gamma = 3 \alpha \omega H,
\end{equation}
which is positive if $\alpha$ is negative. The GCG equation-of-state parameter is given by
\begin{equation}\label{29}
\omega = \frac{p}{\rho} = -\frac{A}{\rho^{\alpha+1}},
\end{equation}
where $A$ is a positive constant. Since $\rho = 3H^2$, we then have, from (\ref{28}),
\begin{equation}\label{30}
\Gamma = -\frac{\alpha A}{3^{\alpha}} H^{-(2\alpha+1)}.
\end{equation}
When $\alpha < 0$ we have energy flux from DE to DM, since $\Gamma > 0$. For $\alpha = 0$ we reobtain the $\Lambda$CDM model with $\Gamma = 0$. For $\alpha = -1/2$ we have a constant $\Gamma$. It was shown in \cite{Wands,GCG} that the split gas presents late-time non-adiabatic perturbations owing to the interaction between the components. This prevents the appearance of oscillations and instabilities present in the power spectrum of the adiabatic GCG \cite{Sandvik,Joras}. Entropic perturbations lead to an effective sound speed equal to zero provided that the DE component is unperturbed \cite{Wands,GCG}. As mentioned above, it was shown in \cite{Zimdahl} that this is true for $\alpha = -1/2$.

\subsection{Quintessence}
\label{Quintessence}

In the FLRW spacetime, the energy density and pressure of a minimally coupled scalar field $\phi$ are given by
\begin{eqnarray}
\rho &=& V + \frac{\dot{\phi}^2}{2}, \label{31} \\
p &=& - V + \frac{\dot{\phi}^2}{2}, \label{32}
\end{eqnarray}
where $V(\phi)$ is the scalar field self-interaction potential. Let us split this field into components
\begin{eqnarray}
\Lambda &=& V - \frac{\dot{\phi}^2}{2}, \quad \; \; \; p_{\Lambda} = -\Lambda, \label{33} \\
\rho_m &=& \dot{\phi}^2, \quad \quad \quad \quad p_m = 0. \label{34}
\end{eqnarray}
The field equations are given, as usual, by
\begin{eqnarray}
3H^2 &=& V + 2H'^2, \label{35} \\
\dot{\phi} &=& -2 H', \label{36}
\end{eqnarray}
where now the prime means derivative with respect to $\phi$. Substituting (\ref{33})-(\ref{34}) into the conservation equation (\ref{8}) we derive the Klein-Gordon equation
\begin{equation}\label{37}
\ddot{\phi} + 3H \dot{\phi} + V'(\phi) = 0.
\end{equation}

All this is general, but we shall now particularise to the special case of a constant $\Gamma$, that is, $\Lambda = 2 \Gamma H$. From (\ref{33}), (\ref{35}) and (\ref{36}) we obtain
\begin{equation}\label{38}
V = \frac{3H^2}{2} + \Gamma H.
\end{equation}
Substituting this potential into (\ref{35}) we have
\begin{equation}\label{39}
4H'^2 + 2\Gamma H - 3H^2 = 0.
\end{equation}
The solution is
\begin{equation}\label{40}
H = \frac{2 \Gamma}{3} \cosh^2 \left( \frac{\sqrt{3}\phi}{4} \right),
\end{equation}
where an integration constant was conveniently chosen. The corresponding potential (\ref{38}) presents a minimum at $\phi = 0$, around which it can be expanded as
\begin{equation} \label{41}
V \approx \frac{4\Gamma^2}{3} + \frac{1}{2} \left( \frac{3\Gamma^2}{4} \right) \phi^2.
\end{equation}
We then see that the mass of the scalar field is given by
$M = \sqrt{3} \Gamma/2$.
On the other hand, at the minimum we have, from (\ref{40}), $H_{dS} = 2\Gamma/3$, a result already derived in Section \ref{Breaking the degeneracy} from (\ref{Borges}).

\section{Conclusions}
\label{Conclusions}

The split GCG and quintessence field described in the last section are examples of a general feature of the dark sector of the Universe, namely the degeneracy between the DM and DE components, related to arbitrariness in the dark energy equation of state. In this letter we have shown that this degeneracy can be broken with the help of two observationally based statements. The first is an appropriate definition of dark matter as the clustering component observed in large scale sctrutures, a definition that is meaningful only if the defined DE component does not cluster. We have seen that this can be achieved with the help of the second statement that dark matter is cold, i.e. non-relativistic, an assumption corroborated by observations. Therefore, we can reduce the degeneracy to two competitive classes of DE models with $\omega = -1$. In particular, this means that an observational joint analysis which identifies DM as the matter observed in large scale structures should naturally give $\omega \approx -1$. At the same time, these results translate the problem of vacuum density to the following question: Is there DM creation at late-time stages of the expansion? If DM cannot be produced at low energies, there is a constant DE density. However, the creation of ultralight and cold DM particles cannot be ruled out. A constant-rate creation would correspond to a DE density which decays linearly with $H$, which would corroborate its association to the vacuum condensate of strongly interacting fields and the corresponding association of DM to condensate fluctuations. Discriminating between these two theoretical possibilities on an observational basis is an exciting endeavour.

${ }$

We are thankful to W. Zimdahl for a critical reading. This work is in memoriam of Prof. Maria Carolina Nemes.

\end{document}